\documentclass[
aps,%
12pt,%
final,%
notitlepage,%
oneside,%
onecolumn,%
nobibnotes,%
nofootinbib,% In the current version of REVTeX, when this option is on,
superscriptaddress,%
noshowpacs,%
centertags]%
{revtex4}
\usepackage{graphicx}
\usepackage{dcolumn}
\usepackage{bm}
\usepackage[none]{hyphenat}

\begin{document}

\title{SHELL-MODEL RESULTS IN
$fp$ AND ${fpg_{9/2}}$ SPACES FOR $^{61,63,65}$Co ISOTOPES}  

%\author{\bf P. C.\ \surname{Srivastava}, \bf V. K. B.\ \surname{Kota}}
%\email{praveen@prl.res.in}
%\author{\bf V. K. B.\ \surname{Kota}}
%\email{vkbkota@prl.res.in}
%\affiliation{\rm Physical Research Laboratory, Ahmedabad, India}

\author{ \bf P.C.~Srivastava}  
\email[]{E-mail: praveen@prl.res.in}
\author{\bf V.K.B.~Kota} 
\email[]{E-mail:vkbkota@prl.res.in} 
\address{\rm Physical Research Laboratory, Ahmedabad, India}

\begin{abstract}   
\vspace{1.0cm}     

Low-lying spectra and several high-spin
states of odd--even $^{61,63,65}$Co isotopes are calculated in two different
shell-model spaces.  First set of calculations have been carried
out in ${fp}$ shell valence space (full $fp$ space for
$^{63,65}$Co and a truncated one for $^{61}$Co) using two
recently  derived ${fp}$ shell interactions, namely GXPF1A
and KB3G, with $^{40}$Ca as core. Similarly, the second set
of calculations  have been performed in ${fpg_{9/2}}$
valence space using an ${fpg}$ effective interaction due
to Sorlin {\it et al}., with $^{48}$Ca as core and
imposing a truncation.  It is seen that
the results of GXPF1A and KB3G are reasonable for
$^{61,63}$Co. For $^{65}$Co, shell-model results show that
the ${fpg}$ interaction adopted in the study is inadequate
and also point out that it is necessary to include
orbitals higher than $1g_{9/2}$ for neutron-rich Co isotopes.

\end{abstract}

%\pacs{21.60.Cs, 27.50.+e} 
\maketitle

%\newpage

\section{Introduction}
\label{s_intro} 

 There are many experimental efforts in the recent years
to study  nuclei around the doubly magic  nucleus
$^{68}$Ni. For example, Lunardi {\it et al.} have recently
populated neutron-rich Fe isotopes from $A=$ 61--66,
through multi-neutron transfer reactions at Legnaro
National Laboratories by coupling the clover detector of
Euroball (CLARA) to a PRISMA magnetic spectrometer
\cite{Lurandi07}. The yrast levels of odd--even
$^{57-60}$Mn isotopes through multi-nucleon transfer
reactions have been reported by Valiente-Dob\'on {\it et al.}
\cite{Dob08}. More recently Steppenbeck  {\it et al.}
\cite{Ste10} have populated high-spin structures in
neutron-rich $^{57-60}$Mn isotopes with fusion--evaporation
reactions induced by  $^{48}$Ca beams at 130 MeV on
$^{13,14}$C targets at Argonne National Laboratory.  Also
for many low-lying levels tentative $J^\pi$assignments
have been made in  \cite{Ste10}. Sorlin {\it et al.}
\cite{Sorlin02} have  populated $^{66,68}$Ni at GANIL and
they proposed evidence for a possible $N=40$ shell gap. In
 \cite{Flanagan09} it is reported that the evolution
of shell structure in $^{75}$Cu isotopes is due to the
inversion of proton 2$p_{3/2}$ and 1$f_{5/2}$  levels. In
another experiment, isomeric low-lying states have been
investigated in $^{75}$Cu \cite{Daugas10}. Using Coulomb
excitation of neutron-rich Zn isotopes, for the first time, 
observation
of the 2$^+$ state in $^{80}$Zn has been reported
in \cite{Walle07}. Theoretically, importance of the
intruder 1$g_{9/2}$ orbital for  lower $fp$ shell nuclei,
namely neutron-rich Cr, Mn, and Fe isotopes, is recently
reported in the literature
\cite{Lurandi07,Dob08,Kaneko08,Sri09,Sria,Srib,Sun09}. In
addition there is  growing experimental evidence for
collective behavior of Cr and Fe isotopes with $N
\sim$ 40 and these results suggest that inclusion of both
1$g_{9/2}$ and  2$d_{5/2}$ orbitals is important for these
nuclei  \cite{Aoi09,Gade10,Ljungvall}. In the
present paper we consider neutron-rich odd Co isotopes.

Experimental information around and beyond $N = 40$ for Co
isotopes is still  limited.  Recently energy levels of
odd--even  $^{61,63,65,67}$Co isotopes have been reported in the
literature. Yrast $\gamma$-ray spectroscopy of neutron
rich $^{61,63}$Co isotopes has been studied by Regan {\it
et al.} \cite{Regan96}. They  have observed a rapid
decrease in observed energy of the yrast $19/2$ state and
yrare  $17/2$ state in $^{63}$Co compared to $^{59}$Co and
$^{61}$Co, possibly giving an indication of the influence
of the neutron $g_{9/2}$ orbital. The level scheme of
$^{65}$Co has been reported for the first time by
Gaudefroy \cite{Gaudefroy05}. Similarly, Pauwels {\it et
al.} \cite{Pauwels08,Pauwels09} have investigated the 
structure of $^{65,67}$Co through $^{65,67}$Fe $\beta$ 
decay and  $^{64}$Ni+$^{238}$U deep-inelastic reaction. 
They have suggested a spherical $7/2^-$ ground state and
also identified the low-lying $1/2^-$ and $3/2^-$ states
in these nuclei. Finally, let us add that Sorlin {\it et
al.} produced $^{67-70}$Co at GANIL \cite{Sorlin00} and
the half-lives for $^{66-71,73}$Co isotopes have been
reported \cite{Mueller99,Weissman99,Mueller00,Sawicka04}.
In the shell-model (SM) framework, calculations for
$^{61,63}$Co isotopes have been carried out by Regan {\it
et al.} \cite{Regan96} using $fp$ valence space  with
$^{48}$Ca core and protons occupying only the $f_{7/2}$
orbit. In this truncated $fp$ space they have used a $fp$
shell interaction developed by van Heese and Glaudemans
\cite{Heese81}. However, there are not yet any shell-model
studies for $^{65,67}$Co isotopes. Following our recent SM
studies for neutron-rich even--even and odd-A isotopes of
Fe \cite{Sri09,Sria}, odd--odd isotopes of Mn \cite{Srib},
and even--even Ni and Zn and odd-A Cu isotopes
\cite{Sri-th},  in the present paper,  we report 
large shell-model calculations for 
$^{61,63,65}$Co isotopes in extended model spaces: (i) full
$fp$ space for $^{63,65}$Co and a truncated $fp$ space
for  $^{61}$Co; (ii) ${fpg_{9/2}}$ space with a truncation
for all isotopes.   The aim of this study is to analyze
the recent experimental data on neutron-rich Co  isotopes
and test the suitability of the chosen valence spaces and
the effective interactions for these isotopes. We have
also made calculations for $^{67}$Co and the results are
discussed briefly at the end of the paper. Now we will
give a preview.

Section 2 gives details of the SM calculations. Sections
3.1--3.3 give the results for odd $^{61}$Co, $^{63}$Co, and $^{65}$Co 
isotopes, respectively. Finally, Section 4 gives conclusions. 

\section {Details of Shell-Model Calculations}

Odd--even $^{61,63,65}$Co isotopes are associated with two shell
closures: $Z = 20$, $N = 20$ and $Z = 20$, $N = 28$. In view of
this, we have performed two sets of calculations. In the first
set, valence space is of full ${fp}$ shell consisting of
1$f_{7/2}$, 2$p_{3/2}$, 1$f_{5/2}$, 2$p_{1/2}$ orbitals for
protons and neutrons and treating $^{40}$Ca as the inert core.
Dimensions of the matrices for $^{61,63,65}$Co isotopes in the
$m$-scheme for $fp$ space are shown in Table 1. In the table
dimensions for $^{67}$Co are also given, as we will discuss the
SM results for this isotope briefly at the end of the paper.
The dimensions for some $J$ values are as large as $\sim 20
\times 10^{6}$ and higher. Note that for $^{61}$Co a
truncation, allowing maximum of four particles (protons plus
neutrons) to be excited from $f_{7/2}$ to the remaining three
$fp$ orbitals, is used in order to make the calculations
tractable. However, the space chosen here is much larger than
that used in the previous shell-model study \cite{Regan96}. 
The $fp$ space calculations are carried out using the recently
derived GXPF1A and KB3G interactions.  The GXPF1 interaction
has been obtained starting from a G-matrix interaction based on
the Bonn-C nucleon--nucleon interaction \cite{Jensen95} and then
modifying the two-body matrix elements by an iterative fitting
calculation to about 700 experimental energy levels in the mass
range $A =$ 47--66 \cite{Honma02}. This interaction predicts a
N=34 subshell gap in $^{54}$Ca and $^{56}$Ti.  However, this
shell closure was not observed in the experimental studies of
the $^{52,54,56}$Ti isotopes \cite{Dinca05b,Fornal04b}.  This
discrepancy led to the modification of the GXPF1 interaction
where  five T=1 two-body matrix elements, mainly involving the
$2p_{1/2}$ and $1f_{5/2}$ orbitals, are adjusted. This modified
interaction is referred as GXPF1A \cite{Honma05}. Similarly,
the KB3G interaction is extracted from the KB3 interaction by
introducing mass dependence and refining its original monopole
changes in order to treat properly the $N=Z=28$ shell closure
and its surroundings \cite{Pov01}. Single-particle energies for
GXPF1A and KB3G interactions  are given in Table 2. 

Second set of calculations have been performed in $fpg_{9/2}$
valence space taking $^{48}$Ca as the inert core.  The 
$fpg_{9/2}$ model space comprises of the  $fp$ ($1f_{7/2}$,
$2p_{3/2}$, $1f_{5/2}$, $2p_{1/2}$) proton orbitals and $rg$
($2p_{3/2}$, $1f_{5/2}$, $2p_{1/2}$, $1g_{9/2}$) neutron
orbitals (with eight $1f_{7/2}$ frozen neutrons). As here the
dimensions of the matrices become very large, a truncation has
been imposed. We used a truncation by allowing up to a total of
four particle excitations from the 1$f_{7/2}$ orbital to the
upper  ${fp}$ orbitals for protons and from the upper $fp$
orbitals to the $1g_{9/2}$ orbital for neutrons.  Dimensions of
the matrices for $^{61,63,65,67}$Co isotopes in the $m$-scheme for
$fpg_{9/2}$ space are shown in Table 1. The dimensions for some
$J$ values are as large as  $\sim 10 \times 10^{6}$ and higher.
For the $fpg_{9/2}$ space, with $^{48}$Ca core, an interaction
reported  by Sorlin {\it et al}. in \cite{Sorlin02} has been
employed. This interaction, called $fpg$ interaction, was built
using $fp$ two-body matrix elements (TBME) from
 \cite{Pov01} and $rg$ TBME from \cite{Nowacki96}. For
the common active orbitals in these subspaces, matrix elements
were taken from  \cite{Nowacki96}. As the latter
interaction ($rg$)  was defined for a $^{56}$Ni core, a scaling
factor of $A^{-1/3}$ was applied to take into account the
change of radius between the $^{40}$Ca and $^{56}$Ni cores. The
remaining $f_{7/2} g_{9/2}$  TBME are taken from  \cite{Kahana69}.
Single-particle energies for the $fpg$
interaction are  given in Table 2.

All the SM calculations have been carried out using the code
{\tt ANTOINE }~\cite{Caurier89,Caurier99} and in this code the
problem of giant matrices is solved by splitting the valence
space into two parts, one for the protons and the other for the
neutrons. The calculations are performed on the 20-node cluster
computer at PRL and the computing time, for example, for
$^{65}$Co is $\sim 8$ days and $\sim 3$ days for $fp$ and
$fpg_{9/2}$ spaces, respectively.

\section{Results and Discussion}

\subsection{$^{\it 61}$Co}

Ground-state spin and parity $7/2^-$ has been assigned for 
$^{61}$Co following the $\beta$ decay of  $^{61}$Fe
\cite{Bron75}. Mateja {\it et al.} \cite{Mateja76} have
identified $9/2^-$ and $11/2^-$ states at excitation energies
of 1285 keV and 1664 keV, using $^{64}$Ni( {\it
p,$\alpha$})$^{61}$Co reaction.  Regan {\it et al.} 
\cite{Regan96}, using the reaction $^{16}$O($^{48}$Ca,{\it
p}2{\it n})$^{61}$Co at a bombarding energy of 110 MeV, were
able to confirm the previously assigned $9/2^-$ and $11/2^-$
states at 1285 keV and 1664 keV, respectively \cite{Mateja76},
and also suggested $1/2^-$  assignment for the 1325-keV level,
which is consistent with the earlier assignment \cite{Coop70}.
The spins and parities of high-spin low-energy levels have been
assigned from angular distributions in reactions and $^{61}$Fe
$\beta$ decay \cite{Mateja76,Regan96}.

In Fig. 1, experimental data are compared with the calculated
energy levels  obtained using SM  with the three different
interactions mentioned before. The calculations predict many
more levels than that are observed experimentally  below 5 MeV,
but we have shown separately only the yrast and yrare levels
that correspond to experimental data (we have not shown the
low-lying $1/2$ and $3/2$ levels and they are discussed
later).  All the interactions predict correct ground-state spin
$7/2^-$. The yrast 9/2$^-$ state is predicted at 1405 keV by
GXPF1A, 1426 keV by KB3G, and 1469 keV by $fpg$ interaction as
compared to the experimental value 1285 keV. The GXPF1A and
KB3G predict the yrast $11/2^-$ above the $9/2^-$ level and the
$fpg$ interaction predict $11/2^-$ level below the $9/2^-$
level. In general the agreement between data and the results
from GXPF1A and KB3G is reasonable for low-lying levels,
as seen from Fig. 1. In
 \cite{Regan96} levels with spins (tentative) $3/2$ and 
$1/2$ are identified at 1028 keV and 1326 keV, respectively. The
calculated excitation energy of the first $3/2^-$ level is 804
keV for GXPF1A, 1242 keV for KB3G, and 745 keV for $fpg$
interactions. Similarly the energies for the first $1/2^-$
level are 1295 keV, 1670 keV, and 1173 keV, respectively.
Therefore the observed levels at 1028 keV and 1326 keV  could
be $3/2^-$ and $1/2^-$ levels predicted by the shell-model.
In addition, experimental high-spin states with $J > 9/2$ look 
more like rotational with nearly regular spacing between levels with
$J$ and $J-1$, while calculations show some irregularity. This shows
that $1g_{9/2}$ orbital is important in $^{61}$Co at high spins. 

Table 3 gives information about wave function structure for
yrast and  few other levels in $^{61}$Co (and also for
$^{63,65}$Co). In this table we have tabulated (i) $S$, sum of
the contributions (intensities)  from particle  partitions
having contribution greater than 1\%;  (ii) $M$, the maximum
contribution from a single partition; and (iii)  $N$, the total
number of partitions contributing to  $S$.  The deviation of 
$S$ from 100\% is due to high configuration mixing. The
increase in  $N$ is also a signature of larger configuration
mixing. From the table one can see that for $^{61}$Co, $S$
varies from $\sim$ 72\% to  $\sim$ 82\%;  the number of
partitions $N$ changing from 13 to 18 and $M$ has variation from
$\sim$ 9\% to $\sim$ 47\%. For the ground state $7/2^-$
predicted by GXPF1A  and KB3G interactions, the partition 
$\pi(0f^{-1}_{7/2})$$\otimes$$\nu(0f^8_{7/2}1p^4_{3/2}0f^2_{5/2})$ 
has intensity 22.5\% and 30\%,  respectively. The calculated
B(E2) values are shown in Table 4 for a few transitions and it
is seen that the B(E2) value for $13/2_1^-$ to $11/2_1^-$  is
quite small compared to all other transitions. However, the
B(E2) for $13/2^-_2$  to $11/2_1^-$ is strong  [B(E2) is $\sim
55$ e$^2$fm$^4$ for the two $fp$ interactions]. It is plausible
that due to the truncation adopted in the  calculations, there
could be a change in the positions of the two $13/2^-$ levels.

\subsection{$^{\it 63}$Co}

Experimentally Runte {\it et al.} \cite{Runte85} have proposed
$7/2^-$ ground state for $^{63}$Co from the $\beta^-$ decay
studies of $^{63}$Fe. Regan {\it et al.} \cite{Regan96}, using
the reaction $^{18}$O($^{48}$Ca,$p2n$)$^{63}$Co at a bombarding
energy of 110 MeV, strongly suggested $J^\pi=9/2^-$ for the
level at 1383 keV, and similarly $11/2^-$ assignment for the
level at 1673 keV has been proposed. In Fig. 2 the calculated
energy levels obtained using SM are compared with experimental
data \cite{Regan96}.  All the three interactions predict
correct ground-state spin. The calculated excitation energy for
the $9/2_1^-$ is 1154 keV for GXPF1A, 1301 keV for KB3G, and
1607 keV for  $fpg$ interaction. Experimentally only one
$9/2^-$ state is observed, but theoretically a second $9/2^-$ 
is predicted at about 1 MeV higher than the first one by all
the interactions (these are not shown in the figure). The first
yrast $3/2^-$ state is predicted by GXPF1A and $fpg$
interactions at $\sim 100$ and $\sim 200$ keV below the
experimental value of 995 keV, while KB3G predicts this level at
$\sim 300$ keV higher.  Similarly, the first $11/2^-$ state is
predicted at 1535 keV by GXPF1A, 1747 keV by KB3G, and 1418 keV
by $fpg$ interaction and the corresponding experimental value
is 1673 keV. 

From Table 3 one can see that for $^{63}$Co, there are 9--17
partitions giving a total of $\sim$ 77--85\% of the total
intensity and the maximum intensity  from a single partition is
$\sim$ 16--45\%. For the $7/2^-$ ground state the partition
$\pi(0f^{-1}_{7/2})$$\otimes$$\nu(0f^8_{7/2}1p^4_{3/2}0f^4_{5/2})$ 
has intensity 31.8\% and 44.8\%,  respectively, for GXPF1A  and
KB3G interactions.   Table 4 gives the calculated  B(E2)
values for a few transitions.  The B(E2) values confirm that
the ground state $7/2^-$ and the  $9/2^-$ (1383 keV) and
$11/2^-$ (2539 keV) levels form a band-like structure. The
$11/2^-$ at 1673 keV belongs to a different structure,
consistent with the claim by Regan {\it et al.} \cite{Regan96}.
The $^{61}$Co
and $^{63}$Co results in Figs. 1 and 2 establish that the $fp$
space is adequate up to about $\sim 3$ MeV excitation for these
nuclei, but the interactions GXPF1A and KB3G require some 
modifications for producing much better agreements with
experimental data. Alternatively, as the
$fpg$ interaction results do not show much improvement over the $fp$ results, a
better $fpg_{9/2}$ interaction may improve the results.

\subsection{$^{\it 65}$Co}

In Fig. 3, the calculated energy levels for $^{65}$Co obtained
using SM with the  three different interactions together with
the experimental data \cite{Pauwels09} are shown. All the
interactions predict correct ground-state spin as observed in
experiment. The experimental energy of the first $9/2^-$ level
is 1479 keV, while the  calculated values are 1753 keV for
GXPF1A, 1943 keV for KB3G, and 2042 keV for $fpg$ interaction.  
Similarly, the first $3/2^-$ state is predicted to be  too low
by GXPF1A and KB3G and at about 250 keV higher by $fpg$
interaction (experimental value is 883 keV).  Above 2 MeV, the
order  of $11/2^-$, $13/2^-$, and $15/2^-$ is seen in experimental
data is different from the order given by
GXPF1A and KB3G interactions. However, the $fpg$ interaction  gives
$11/2^-$ as lowest and then nearly degenerate $13/2^-$ and
$15/2^-$ levels. Lowest $3/2^-$ and $11/2^-$ are much lower for
GXPF1A and KB3G interactions and their positions are better
predicted by the $fpg$ interaction. Combining all these, we
conclude that for $^{65}$Co,  $fpg$ interaction gives better results.
However the energies predicted by the
$fpg$ interaction show that modifications of this interaction
are clearly needed. 

Table 3 gives some information about wave function structure
as generated by the $fpg$ interaction for $^{65}$Co. It is seen
that $\sim$ 11--18 partitions  generate $\sim$ 81--84\%  of the
intensity. The calculated B(E2) values for $fpg$ interaction
for 9/2$_1^-$ $\rightarrow$ 7/2$_1^-$, 11/2$_1^-$ $\rightarrow$
9/2$_1^-$,\\ 11/2$_2^-$ $\rightarrow$ 9/2$_1^-$  transitions are
105.12, 0.28, and 1.29 $e^2$fm$^4$, respectively. Thus, there is
a change of structure  after $9/2^-$ with $g_{9/2}$ occupancy
for neutrons becoming large. For $11/2_1^-$ the occupancy is
$\sim$ 0.8 and for  $11/2_2^-$ it is $\sim$ 1.5. Similarly,
wave functions show that the $11/2_1^-$ states contain partitions
with $g_{9/2}$ orbital contributing to $\sim$ 40--60\% of the
intensity.

For further tests of the adopted $fpg$ interaction, we have
made SM calculations also for $^{67}$Co. Its ground-state spin
$7/2^-$ is correctly predicted  by the SM calculations. 
Pauwels {\it et al.} \cite{Pauwels08} have identified (1/2$^-$)
isomeric state in $^{67}$Co with a half-life of 496(33) ms at an
unexpected low energy of 492 keV.  It is claimed \cite{Heyde83}
that strong proton--neutron correlations  inducing deformation
are responsible for the decrease in excitation energy of  the
1/2$^-$ state.  Firstly, the position of the first $1/2^-$ and
$3/2^-$ (680 keV) levels are predicted to be at very high
excitation ($> 4$ MeV) compared to data by GXPF1A and KB3G
interactions. Thus clearly $fp$ space is not adequate for this
nucleus. The SM calculations with $fpg$ interaction gave the
first $1/2^-$, $3/2^-$, and $5/2^-$ at 2118 keV, 1157 keV, and
1728 keV, respectively, while the data values are  492 keV, 680
keV, and 1252 keV. Thus, for a good description of the spectra
of neutron-rich Co isotopes, the $fpg$ interaction
adopted in the present study clearly needs modifications.
Towards this end we have made calculations for $^{65}$Co using
various monopole corrections \cite{Cau-05}.

A significant decrease of the energy gap between proton
1$f_{7/2}$ and 1$f_{5/2}$ orbitals is obtained for the $fpg$
interaction with the increase of the occupation of the $\nu
g_{9/2}$ orbital, usually referred to as monopole migration.
Otsuka {\it et al.} in \cite{Otsuka06} have shown that
this is due to opposite  actions of the strong proton--neutron
tensor force on the $\pi f_{7/2}$ and  $\pi f_{5/2}$ orbitals
while filling the neutrons in the 1$g_{9/2}$ orbital.  To study
the importance of the monopole corrections to the $fpg$
interaction, we have modified the $1g_{9/2}1f_{7/2}$ matrix
elements by subtracting 100 keV and $g_{9/2}f_{5/2}$ matrix
elements by adding 100 keV, so that the $1f_{7/2}$--$1f_{5/2}$
proton gap is increased  for neutron-rich nuclei. With this
modified interaction, the $3/2_1^-$, $11/2_1^-$, $9/2_1^-$,  and
$1/2_1^-$ levels in $^{65}$Co are predicted at 740, 1115, 1584,
and 1880 keV,  respectively. The $1/2_1^-$ level is still
predicted at much higher energy  than the  experimental value
and also the  $11/2_1^-$ level is too low. Calculations are
also performed by changing the original matrix elements by  200
keV and found no significant improvement in the results. In
addition, calculations are also carried out by modifying
$p_{1/2}g_{9/2}$  matrix elements (by 50 -- 200 keV) and also
$p_{3/2}g_{9/2}$ matrix elements (by 100 keV).  Again, not much
improvement is seen in the above results.  All these results clearly
show that one has to  go beyond $fpg_{9/2}$ space and include
higher orbitals.

\section{Conclusions}

In the present work, the results of large scale
shell-model calculations are reported for neutron-rich
odd--even isotopes of Co with $A=61$, 63, and 65 in two valence
spaces: full ${fp}$ space and $fpg_{9/2}$ space with $^{48}$Ca
core. For $^{61}$Co a truncation has been adopted, but the
matrix dimensions for the present calculations are larger
than the previous $fp$ calculations.  On the other hand, full
$fp$ space results with the recently derived GXPF1A and KB3G
interactions are reported for $^{63,65}$Co. Similarly, for the 
${fpg_{9/2}}$ space calculations, a truncation has been adopted
and the so-called a $fpg$ interaction is employed. Results for
energy spectra shown in Figs. 1 and 2 for $^{61,63}$Co confirm
(see also the discussion in Sections 3.1 and 3.2) that the $fp$
space results with GXPF1A and KB3G are reasonable when 
compared with data (note that in \cite{Regan96} shell-model
results are  presented only in a restricted $fp$ space and also
with a special effective  interaction). For $^{65}$Co the
results in the extended model space, i.e., in ${fpg_{9/2}}$ space
do not reproduce correct experimental findings, and further,
with monopole correction by reducing the gap between
$f_{7/2}$--$f_{5/2}$, the 1/2$_1^-$ is still high and 11/2$_1^-$
is low. Thus it appears to be necessary to include the
intruder 2$d_{5/2}$ orbital  while approaching  $N\sim 40$.

In the last decade good $fp$ space effective interactions, with
$^{40}$Ca core, have been generated giving good SM spectroscopy
\cite{Cau-05}. Recently, a good effective interaction (JUN45)
in $f_{5/2}pg_{9/2}$ space, with $^{56}$Ni core, has been
reported with an extensive set of SM calculations by  Honma
{\it et al.} \cite{Honma09}. In addition, the so-called $fpg$
interaction in $fpg_{9/2}$ space with $^{48}$Ca core has some
limited success \cite{Sorlin02,Srib}. However, combining the results
of the present paper with our previous large shell-model studies of 
neutron-rich
Fe and Mn isotopes \cite{Sri09,Sria,Srib}, we conclude that it
is necessary to generate good  $fpg_{9/2}$ and $fpg_{9/2}d_{5/2}$ space
effective interactions (with $^{48}$Ca core)  for neutron-rich Fe, Mn,
and Co isotopes. After the present work was complete, there appeared a report
giving some results for low-lying  
yrast levels in even--even 
Cr and Fe isotopes, showing collective effects as N 
approached 40,
obtained using a new hybrid interaction in
$fpg_{9/2}d_{5/2}$ space \cite{Now-pre}. It remains to be seen if this
interaction can describe (here matrix
dimensions will be $\sim 10^{10}$ and higher) the experimental data on 
neutron-rich odd Co isotopes.
  
One of the authors (PCS)  would like to thank  E.~Caurier and
F.~Nowacki  for providing the shell-model code ANTOINE and 
F.~Nowacki for supplying $fpg$ interaction matrix elements.
His special thanks are also due to P.~Van Isacker, M.~Rejumund, and
I.~Mehrotra for their help from time to time.

\newpage

{\bf Table 1.} Dimensions of the shell-model matrices for  $^{61,63,65,67}$Co  
in the $m$-scheme for $fp$ and $fpg_{9/2}$ spaces. The $fp$ space dimensions given in the
table for $^{63}$Co, $^{65}$Co and $^{67}$Co are for the full $fp$ space. However, for $^{61}$Co,
the dimensions are given for the truncated $fp$ space with maximum of four particles 
excited from 1$f_{7/2}$ orbital to the rest of the $fp$ orbitals. 
For $fpg_{9/2}$ space, the dimensions are for the truncation explained in the text.\\

{\bf Table 2.} Single-particle energies (in MeV) for GXPF1A, KB3G, and $fpg$ interactions.\\

{\bf Table 3.} The extent of configuration mixing involved in $^{61,63,65}$Co
isotopes for different states (for each state the numbers quoted are $S$, 
sum  of the contributions  from  particle
partitions,  each  of  which  is contributing greater than 1\%;
 $M$, maximum contribution from  a  single  partition; and
$N$, total number of partitions contributing to $S$ ). Note
that $S$ and $M$ are in percentage.\\

{\bf Table 4.} Calculated B(E2) values for some transition for
 $^{61,63}$Co isotopes with standard effective charges:
 e$_{\rm eff}^\pi$=1.5$e$, e$_{\rm eff}^\nu$=0.5$e$.  All B(E2) values are
 in $e$$^2$fm$^4$ unit (the experimental $\gamma$-ray energies
 corresponding to these transitions are also shown).\\
 
\newpage

{ \bf Fig. 1.} Experimental data \cite{Regan96} for $^{61}$Co compared
with shell-model results generated by three different effective interactions.
See text for details.\\

{\bf Fig. 2.} The same as in Fig.1, but for $^{63}$Co. \\

{\bf Fig. 3.} The same as in Fig.1, but for $^{65}$Co. \\

\newpage

\begin{table}
\setcaptionmargin{5mm}
\onelinecaptionsfalse
\captionstyle{flushleft}
\caption{ Dimensions of the shell-model matrices for  $^{61,63,65,67}$Co  
in the $m$-scheme for $fp$ and $fpg_{9/2}$ spaces. The $fp$ space dimensions given in the
table for $^{63}$Co, $^{65}$Co and $^{67}$Co are for the full $fp$ space. However, for $^{61}$Co,
 the 
dimensions are given for the truncated $fp$ space with maximum of four particles 
excited from 1$f_{7/2}$ orbital to the rest of the $fp$ orbitals. 
For $fpg_{9/2}$ space, the dimensions are for the truncation explained in the text.}
\label{t_dim}
\begin{center}
%\resizebox{7.5cm}{!}{
%\resizebox{!}{7.5cm}{
\begin{tabular}{c|c|c|c|c|c|c|c|c}
\hline
 \multicolumn{5}{c|}{$fp$ space($^{40}$Ca core)}   & \multicolumn{4}{|c}{$fpg_{9/2}$ space($^{48}$Ca core)}\cr
\hline
% &  \multicolumn{2}{c|} \\ 
 2J$^\pi$ &~$^{61}$Co &~~~~~ $^{63}$Co&~~~~~ $^{65}$Co&~~~~ $^{67}$Co&~~~~ $^{61}$Co&~~~~~ $^{63}$Co&~~~~~ $^{65}$Co&~~~~ $^{67}$Co\cr
\hline
1$^-$ &~ 17034417& ~ 28464525&~ 1227767& ~ 7531 &~ 9480566& ~12957305& ~ 6308001&~  874872  \cr
\hline
3$^-$ &~ 16458582& ~ 27480719&~ 1176872& ~ 7116 &~ 9188151& ~12599047& ~ 6139881& ~ 850942  \cr
\hline
5$^-$ &~ 15361866& ~ 25609431&~ 1080985& ~ 6352 &~ 8629526& ~11911108& ~ 5816312& ~ 804889  \cr
\hline
7$^-$ &~ 13846120& ~ 23028746&~~ 950895& ~ 5347 &~ 7853482& ~10947190& ~ 5361260& ~ 740166  \cr
\hline
9$^-$ &~ 12044787& ~ 19971189&~~ 800263& ~ 4226 &~ 6924099& ~ 9778867& ~ 4806873& ~ 661360  \cr
\hline
11$^-$&~ 10105027& ~ 16691144&~~ 643552& ~ 3135 &~ 5912846& ~ 8487672& ~ 4190319& ~ 573863  \cr
\hline
13$^-$&~~  8167820& ~ 13430682&~~ 493609&  ~~2161 &~~ 4888828& ~ 7155066& ~ 3549245& ~ 483060 \cr 
\hline
15$^-$&~~  6353114& ~ 10392710&~~ 360351&  ~~1378 &~~ 3912382& ~ 5855328& ~ 2918851& ~ 394088  \cr
\hline 
17$^-$&~~  4747943& ~~ 7721925& ~~249638& ~~~ 801 &~~ 3028771& ~ 4648184& ~ 2328191& ~ 311094  \cr 
\hline
19$^-$&~~  3403216& ~~ 5499474& ~~163574& ~~~ 421 &~~ 2267035& ~ 3576504& ~ 1799122& ~ 237276  \cr 
\hline
\end{tabular}
\end{center}
\end{table}

\newpage
\begin{table}
\setcaptionmargin{5mm}
\onelinecaptionsfalse
\captionstyle{flushleft}
\caption{ Single-particle energies (in MeV) for GXPF1A, KB3G, and $fpg$ interactions.}
\label{t_spe}
\begin{center}
%\resizebox{7.5cm}{4.5cm}{
\begin{tabular}{c|c|c|c}
\hline
Orbital  & ~~GXPF1A &~~ KB3G &~~ $fpg$ \\		   
\hline
 1$f_{7/2}$&~~ -8.6240 &~~  0.0000   &~~ 0.000   \\
\hline
 2$p_{3/2}$&~~ -5.6793 &~~ 2.0000   &~~ 2.000   \\
\hline
 2$p_{1/2}$&~~ -4.1370 &~~ 4.0000   &~~ 4.000   \\
\hline
 1$f_{5/2}$&~~ -1.3829 &~~ 6.5000   &~~ 6.500   \\
\hline
 1$g_{9/2}$&~~    -    &~~  -       &~~ 9.000   \\
\hline            
\end{tabular}
\end{center}
\end{table}

\newpage
\begin{table} 
\setcaptionmargin{5mm}
\onelinecaptionsfalse \captionstyle{flushleft} \caption{ 
The extent of configuration mixing involved in $^{61,63,65}$Co
isotopes for different states (for each state the numbers quoted are $S$, 
sum  of the contributions  from  particle
partitions,  each  of  which  is contributing greater than 1\%;
 $M$, maximum contribution from  a  single  partition; and
$N$, total number of partitions contributing to $S$ ). Note
that $S$ and $M$ are in percentage.} 
\begin{center}
%\resizebox{7.5cm}{!}{ %\resizebox{!}{7.5cm}{
\begin{tabular}{c|c|c|c|c|c|c|c|c|c|c|c} 
\hline
% $J^\pi$&$^{61}_{27}$Co$_{34}$
$J^\pi$& \multicolumn{3}{|c|}{$^{63}_{27}$Co$_{36}$}&$J^\pi$&\multicolumn{3}{|c|}{$^{65}_{27}$Co$_{38}$}&$J^\pi$&\multicolumn{3}{|c}{$^{65}_{27}$Co$_{38}$}\cr
\cline{2-4}\cline{6-8}\cline{10-12}
       &$S$&~ $M$&~ $N$   &         &$S$&~ $M$&~ $N$   &         &$S$~ &$M$~ &$N$       \cr
\cline{2-4}\cline{6-8}\cline{10-12}
&\multicolumn{3}{|c|}{GXPF1A} &&\multicolumn{3}{|c|}{GXPF1A}&&\multicolumn{3}{|c}{$fpg$}\\
\hline
$7/2_1^-$ & ~71.6 & ~22.5 & ~18   & $7/2_1^-$  & 80.7 & ~31.8 & ~14 & $7/2_1^-$ &~83.0 & ~47.9& ~11\cr
\hline
$3/2_1^-$&~63.4& ~9.0&~18   & $3/2_1^-$  &~~76.8& ~16.1& ~15& $3/2_1^-$&~81.1& ~27.2& ~16\cr
\hline
$9/2_1^-$&~71.9& ~16.2& ~17   & $9/2_1^-$  &~~85.1& ~34.1& ~17&$9/2_1^-$&~84.3& ~42.9& ~16 \cr
\hline
$11/2_1^-$&~67.4& ~15.5& ~17  & $11/2_{1}^-$&~~80.8& ~28.6& ~15&$11/2_{1}^-$&~83.0& ~26.9& ~17\cr
\hline
$13/2_{1}^-$&~74.4& ~35.0& ~12 &$11/2_{2}^-$&~~80.7& ~30.2&~14&$11/2_{2}^-$&~78.1& ~12.1&~18\cr
\hline
$13/2_{2}^-$&~75.5& ~30.7& ~15 &$13/2_{2}^-$ &~~81.6& ~30.5& ~13\cr
\hline

&\multicolumn{2}{c}{KB3G}&&&\multicolumn{4}{c}{KB3G}~~\\
\cline{1-8}
$7/2_1^-$&~76.6& ~30.0& ~18   &$7/2_1^-$&~~82.9& ~44.8& ~12  \cr
\cline{1-8}
$3/2_1^-$&~71.1& ~20.8& ~14      &$3/2_1^-$&~~82.9& ~30.3& ~15\cr
\cline{1-8}
$9/2_1^-$&~75.5& ~13.0& ~16      &$9/2_1^-$&~~81.9& ~31.6& ~11\cr
\cline{1-8}
$11/2_1^-$&~76.3& ~27.4& ~17     &$11/2_{1}^-$&~~81.1& ~35.7& ~11\cr
\cline{1-8}
$13/2_{1}^-$&~82.0& ~47.5& ~13 &$11/2_{2}^-$&~~83.1& ~32.8& ~11\cr
\cline{1-8}
$13/2_{2}^-$&~79.8& ~37.7& ~15 &$13/2_{2}^-$ &~~82.5& ~35.8&~~9\cr
\hline
\end{tabular}
\end{center}
\end{table}

\newpage
\begin{table}
\setcaptionmargin{5mm}
\onelinecaptionsfalse
\captionstyle{flushleft}
\caption{Calculated B(E2) values for some transition for
 $^{61,63}$Co isotopes with standard effective charges:
 e$_{\rm eff}^\pi$=1.5$e$, e$_{\rm eff}^\nu$=0.5$e$.  All B(E2) values are
 in $e$$^2$fm$^4$ unit (the experimental $\gamma$-ray energies
 corresponding to these transitions are also shown).}
\begin{center}
%\resizebox{8.5cm}{!}{
\begin{tabular}{c|c|c|c|c}
\hline
Nucleus & Transition & $E_\gamma$, keV &~  GXPF1A &~~ KB3G   \\		   
\hline
$^{61}$Co~~~ &$B(E2$; ~9/2$_1^-$ $\rightarrow$ ~7/2$_1^-$  ) &~~ 1285&~~ 209.47 &~~ 132.60  \\
\cline{2-5}
             &$B(E2$; 11/2$_1^-$ $\rightarrow$ ~9/2$_1^-$  ) &~~ 379&~~ 144.61 &~~ 31.20  \\
\cline{2-5}
             &$B(E2$; 13/2$_1^-$ $\rightarrow$ 11/2$_1^-$  ) &~~ 710&~~ 8.13 &~~ 0.54  \\
\cline{2-5}
             &$B(E2$; 15/2$_1^-$ $\rightarrow$ 13/2$_1^-$  ) &~~ 752&~~ 31.18 &~~ 65.50  \\
\hline
$^{63}$Co~~~ &$B(E2$; ~9/2$_1^-$ $\rightarrow$ ~7/2$_1^-$  ) &~~ 1383&~~ 164.55 &~~ 131.84 \\
\cline{2-5}
             &$B(E2$; 11/2$_2^-$ $\rightarrow$ ~9/2$_1^-$) &~~ 1156&~~ 157.78 &~~ 120.35 \\
\cline{2-5}
             &$B(E2$; 11/2$_1^-$ $\rightarrow$ ~9/2$_1^-$) &~~ 290&~~ 2.04 &~~ 2.92 \\
\cline{2-5}
             &$B(E2$; 13/2$_1^-$ $\rightarrow$ 11/2$_1^-$  ) &~~ 495&~~  42.44 &~~ 60.50  \\

\hline           
\end{tabular}
\end{center}
\end{table}

\pagebreak

\newpage

\vspace{4.0cm}
\begin{figure}
\setcaptionmargin{5mm}
\onelinecaptionsfalse
\includegraphics[width=16.4cm]{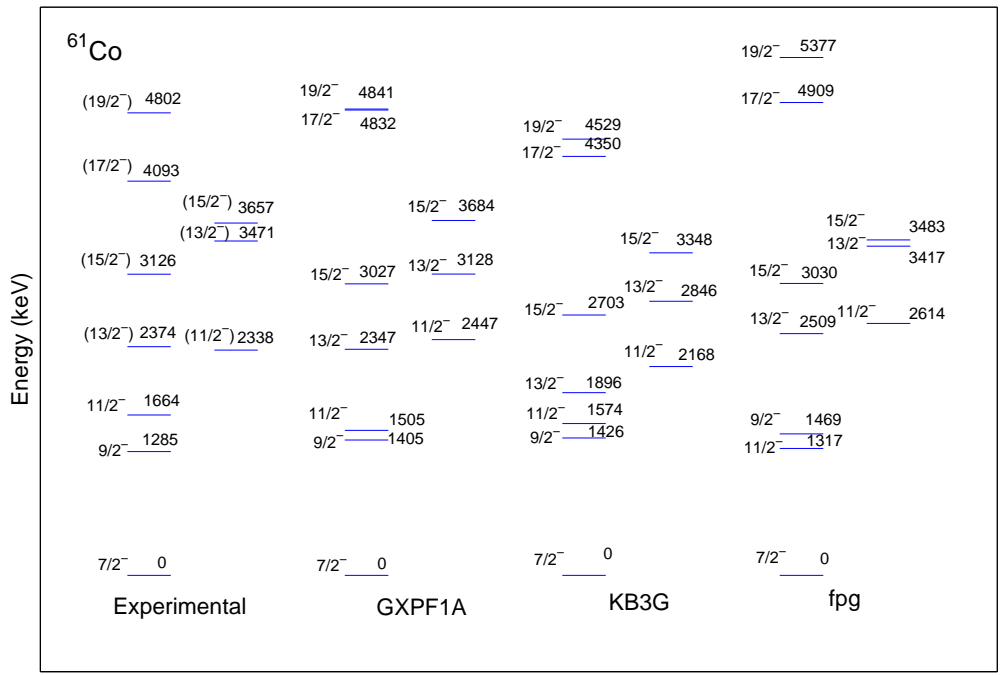}
\captionstyle{normal} 
\caption{
Experimental data \cite{Regan96} for $^{61}$Co compared
with shell-model results generated by three different effective interactions.
See text for details.}
\label{f_co61}
\end{figure}

\pagebreak

\begin{figure}
\setcaptionmargin{5mm}
\onelinecaptionsfalse
\includegraphics[width=16cm]{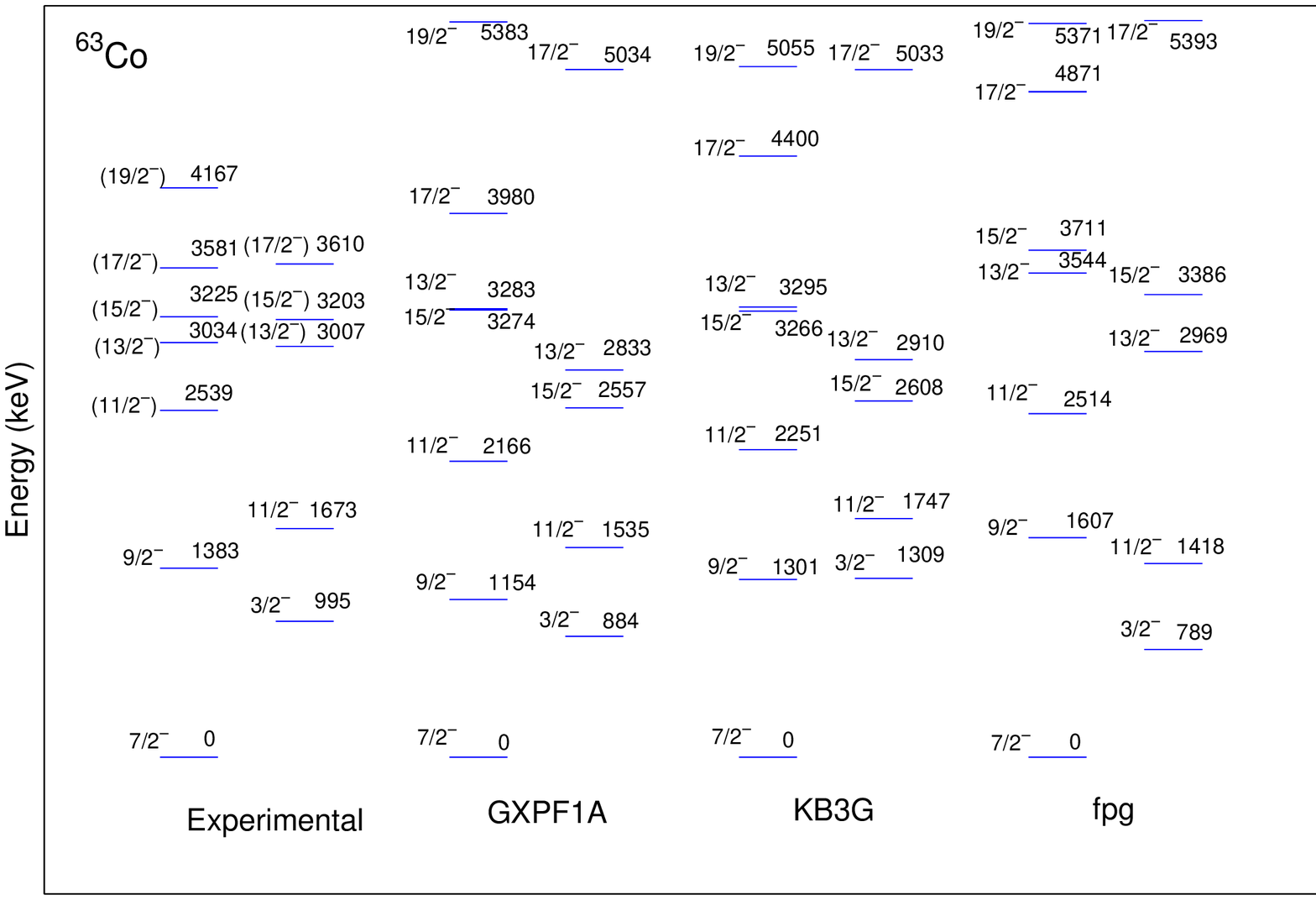}
\captionstyle{normal} 
\caption{
The same as in Fig.1, but for $^{63}$Co.}
\label{f_co63}
\end{figure}

\newpage

\vspace{4.0cm}
\begin{figure}
\setcaptionmargin{5mm}
\onelinecaptionsfalse
\includegraphics[width=16cm]{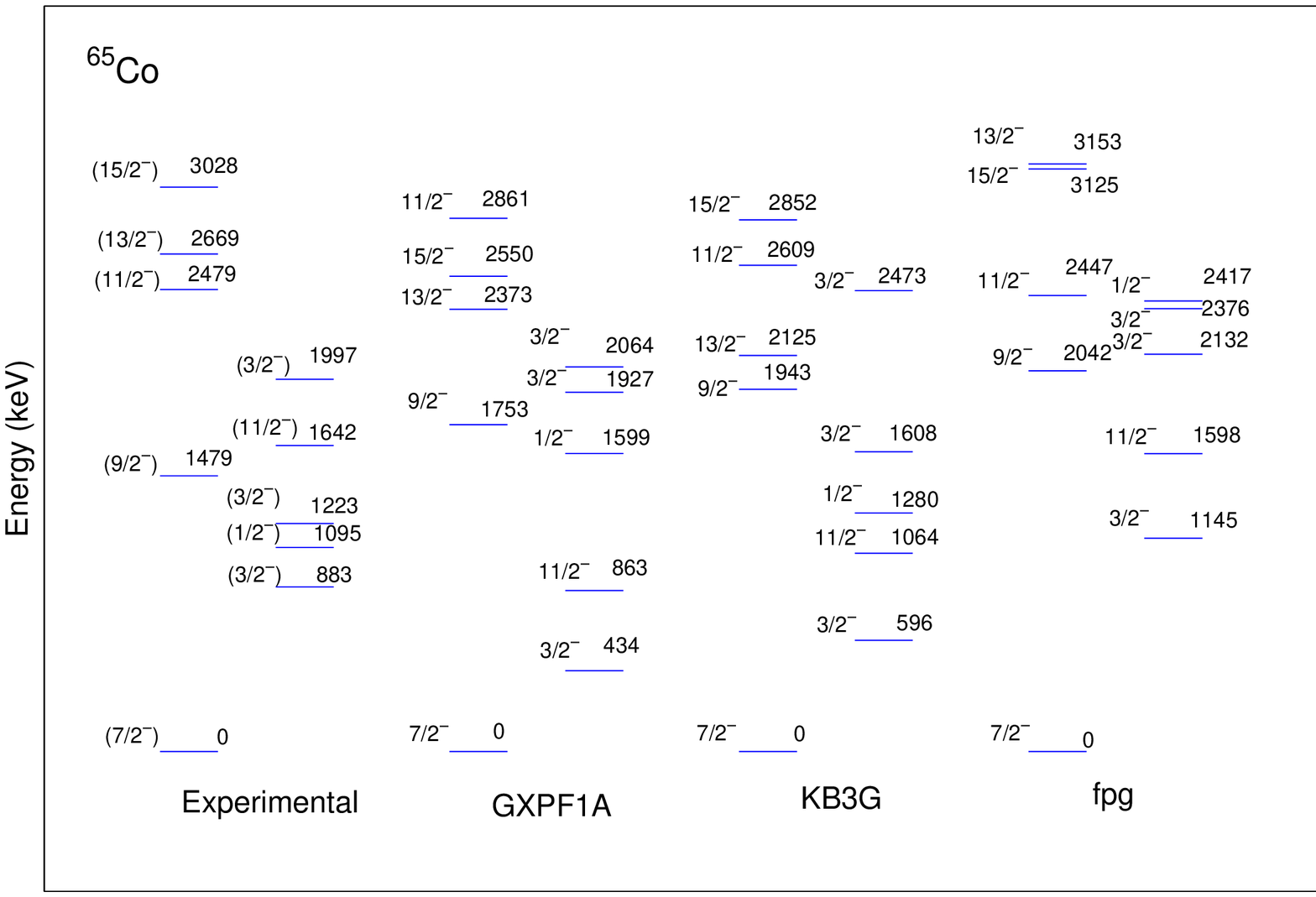}
\captionstyle{normal} 
\caption{
The same as in Fig.1, but for $^{65}$Co.}
\label{f_co65}
\end{figure}


\begin{thebibliography}{99}

\bibitem{Lurandi07}
S.~Lunardi {\it et al.},
Phys.\ Rev.\ C {\bf76},  034303 (2007).

\bibitem{Dob08}
J. J.~Valiente-Dob\'on {\it et al.},
Phys.\ Rev.\ C {\bf78}, 024302 (2008).

\bibitem{Ste10}
D.~Steppenbeck {\it et al.},
Phys.\ Rev.\ C {\bf81}, 014305 (2010).

\bibitem{Sorlin02}
O.~Sorlin {\it et al.},
Phys.\ Rev.\ Lett. {\bf88}, 092501 (2002).


\bibitem{Flanagan09}
K. T. ~Flanagan {\it et al.},
Phys.\ Rev.\ Lett. {\bf103},  142501 (2009).

\bibitem{Daugas10}
J. M.~Daugas {\it et al.},
Phys.\ Rev.\ C {\bf81},  034304 (2010).

\bibitem{Walle07}
J. Van de ~Walle {\it et al.},
Phys.\ Rev.\ Lett. {\bf99},  142501 (2007).

\bibitem{Kaneko08}
K.~Kaneko, Y. Sun, M.~Hasegawa, and T.~Mizusaki,
Phys.\ Rev.\ C {\bf78}, 064312 (2008).


\bibitem{Sri09}
P. C.~Srivastava and I.~Mehrotra,
J.\ Phys.\ G\ {\bf36},  105106 (2009).

\bibitem{Sria}
P. C.~Srivastava and I.~Mehrotra, 
Phys.\ Atom.\ Nucl.\ {\bf73}, 1656 (2010).

\bibitem{Srib}
P. C.~Srivastava and I.~Mehrotra,
Eur.\ Phys.\ J.\  A\ {\bf45},  185 (2010).

\bibitem{Sun09}
Y. Sun, Y.-C. Yang, H.-L. Liu, K. Kaneko, M. Hasegawa, and T. Mizusaki,
Phys.\ Rev.\ C {\bf80}, 054306 (2009).

\bibitem{Aoi09}
N. Aoi {\it et al.},
Phys.\ Rev.\ Lett. {\bf102}, 012502 (2009).

\bibitem{Gade10}
A.~Gade {\it et al.},
Phys.\ Rev.\ C {\bf81},  051304(R) (2010).

\bibitem{Ljungvall}
J.~Ljungvall {\it et al.},
Phys.\ Rev.\ C {\bf81},  061301(R) (2010).

\bibitem{Regan96} P. H.~Regan, J. W.~Arrison, U. J.~H\"{u}ttmeier,
and D.P.~Balamuth, Phys.\ Rev.\ C {\bf54},  1084 (1996).

\bibitem{Gaudefroy05}
L.~Gaudefroy,
Ph.D. Thesis, Universit\'e de Paris XI, Orsay (2005).

\bibitem{Pauwels08}
D. ~Pauwels {\it et al.},
Phys.\ Rev.\ C {\bf78},  041307(R) (2008).

\bibitem{Pauwels09}
D. ~Pauwels {\it et al.},
Phys.\ Rev.\ C {\bf79},  044309 (2009).


\bibitem{Sorlin00}
O.~Sorlin {\it et al.},
Nucl.\ Phys. A {\bf 669},   351 (2000).

\bibitem{Mueller99}
W. F.~Mueller {\it et al.},
Phys.\ Rev.\ Lett. {\bf83},   3613 (1999).

\bibitem{Weissman99}
L.~Weissman {\it et al.},
Phys.\ Rev.\ C {\bf59},  2004 (1999).

\bibitem{Mueller00}
W. F.~Mueller {\it et al.},
Phys.\ Rev.\ C {\bf61},  054308 (2000).

\bibitem{Sawicka04}
M.~Sawicka {\it et al.},
Eur.\ Phys.\ J.\ A {\bf22}, 455 (2004).

\bibitem{Heese81}
A. G. M. van ~Heese and P. W. M. Glaudemans,
Z.\ Phys. \ A  {\bf303},  267 (1981).

\bibitem{Sri-th} P. C.~Srivastava, Ph.D. Thesis, University of
Allahabad (Allahabad, India, 2010).

\bibitem{Jensen95}
M. Hjorth-Jensen, T.T.S. Kuo, and E. Osnes,
Phys.\ Rept. {\bf261},   125 (1995).

\bibitem{Honma02}
M.~Honma, T.~Otsuka, B. A.~Brown, and T.~Mizusaki, 
Phys.\ Rev.\ C {\bf65},  061301(R) (2002).

\bibitem{Dinca05b}
D. -C.~Dinca {\it et al.},
Phys.\ Rev.\ C {\bf71},  041302(R) (2005).

\bibitem{Fornal04b}
B.~Fornal {\it et al.},
Phys.\ Rev.\ C {\bf70},   064304 (2004).

\bibitem{Honma05}
M.~Honma,T.~Otsuka, B.A.~Brown, and T.~Mizusaki, 
Eur.\ Phys.\ J.\ A {\bf25}, 499 (2005).

\bibitem{Pov01}
A.~Poves, J. Sanchez-~Solano, E.~Caurier, and F. Nowacki,
Nucl.\ Phys. A {\bf 694},  157 (2001).

\bibitem{Nowacki96}
F.~Nowacki, Ph.D. Thesis
(IRes, Strasbourg, 1996).

\bibitem{Kahana69}
S.~Kahana, H.C.~Lee, and C.K.~Scott,
Phys.\ Rev.\  {\bf180},  956 (1969).

\bibitem{Caurier89}
E.~Caurier, code {\tt ANTOINE}
(Strasbourg, 1989), unpublished.

\bibitem{Caurier99}
E.~Caurier and F.~Nowacki,
Acta Phys.\ Pol.\ B {\bf30}, 705 (1999).


\bibitem{Bron75}
J.~Bron, H.W.~Jongsma, and H.~Verhul,
Phys.\ Rev.\ C {\bf11}, 996 (1975).

\bibitem{Mateja76}
J.F.~Mateja {\it et al.},
Phys.\ Rev.\ C {\bf13}, 2269 (1976).

\bibitem{Coop70}
K.L.~Coop, I.G.~Graham, and E.~Titterton,
Nucl.\ Phys.\ A {\bf 150}, 346 (1970).

\bibitem{Runte85}
E.~Runte {\it et al.},
Nucl.\ Phys. A {\bf 441}, 237 (1985).

\bibitem{Heyde83}
K.~Heyde, P.~Van~Isacker, M.~Waroquier, J.L.~Wood, and R.A.~Mayer,
Phys.\ Rept. {\bf102},  291 (1983).

\bibitem{Cau-05} E.~Caurier, G.~Martinez-Pinedo, F.~Nowacki, 
A.~Poves, and A.P.~Zuker, Rev. Mod. Phys. {\bf 77}, 427 (2005).

\bibitem{Otsuka06}
T. Otsuka, T. Matsuo, and D. Abe,
Phys.\ Rev.\ Lett. {\bf97},  162501 (2006).
 
\bibitem{Honma09}
M.~Honma, T.~Otsuka, T.~Mizusaki, and M.~Hjorth-Jensen, 
Phys.\ Rev.\ C {\bf80}, 064323 (2009).

\bibitem{Now-pre}
S.M.~Lenzi, F. Nowacki, A. Poves, and K. Sieja, Phys.\ Rev.\ C {\bf82}, 054301 (2010).

\end{thebibliography}
\end{document}